\documentstyle[prb,preprint,aps]{revtex}

\begin{document}
\input{psfig}

\title
{\bf Quantum Computation with Aharonov-Bohm Qubits}

\author{ A. Barone$^1$, T. Hakio\u{g}lu$^2$ and I. O. Kulik$^2$}
\address{
$^1$Department of Physical Sciences, University of Naples Federico II,
P. Tecchio 80, Naples 80125, Italy\\
$^2$Department of Physics, Bilkent University, Ankara 06533, Turkey\\}

\maketitle
\begin{abstract}
We analyze the possibility of employing a mesoscopic/nanoscopic ring
of a normal metal in a double-degenerate persistent-current state {\bf with
 a third auxhiliary level and} in
the presence of the Aharonov-Bohm flux equal to the half of the
normal flux quantum, $hc/e$, as a qubit.
The auxhiliary level can be effectively used for all fundamental
quantum logic gate (qu-gate) operations
which includes the initialization, phase rotation,
bit-flip, and the Hadamard transformation, as well as the
double-qubit controlled operations (conditional bit flip).
We suggest a tentative realization of the mechanism proposed
as either the mesoscopic structure of three quantum dots
coherently coupled by resonant tunneling in crossed magnetic and
electric fields, or as a nanoscopic structure of triple anionic
vacancy (similar to $F_3$ centers in alkali halides) with one
trapped electron in one spin projection state.
\end{abstract}


\vspace{5mm}
\newpage

\section{Introduction} Quantum computation [1-3] is a promising
tool for solving intractable mathematical problems, those in which
the number of computational steps (if solved with a classical
computer) increases exponentially with the number of computational
units ($N$), e.g., number of spins in the Heisenberg
ferromagnet, number of electrons and lattice sites in the Hubbard
model of solid, number of binary digits in a large integer to be
factorized, etc. If these units (spins, atoms, digits) are
represented as ``quantum bits'' [4] and processed by unitary
transformations acted upon by the logical quantum gates [5], at
least some of these problems can be solved in a polynomial time in
$N$ (e.g., the Shor's algorithm [6] for factorizing large
integers). Basically, the fundamental gates are unitary time evolutions for
given Hamiltonians executed on qubits or on pairs of qubits and for
certain time intervals. Fundamental gates are known to be the
unitary operations such as the bit-flip, phase-flip, Hadamard and
the controlled-NOT (CNOT) operations.
Workers in the field at earlier times considered qubit
realizations as quantum optical or atomic systems, and shifted at
more recent times to other methods employing mesoscopic condensed
matter structures (quantum dots [7], superconducting Cooper-pair
boxes [8-11], Josephson junctions [12-14]) as qubits). In
Ref.[15], the necessary conditions for quantum computation have
been specified, not all of which have already achieved perfect
realization (the problems with the solid-state qubits are
documented in Refs.[16,17]). This leaves space for more
suggestions of the instrumental realization of qubits, especially
those that use the solid state technology. We investigate in this
paper one of such possibilities. The particular advantage of the suggested
three state mesoscopic persistent current system is the fact that
no physical coupling or switch is required in the manifestation in any
of the fundamental quantum gates. (Similar ideas have been suggested
earlier [18], however without any clear instrumental realization.)
In addition, the auxhiliary level can also be used to coherently
couple the operational qubit states to the environment including the
other qu-gates as well as the input-output devices, and hence it is expected
to help also in the manifestation of
possible error correction mechanisms. The proposed structure is naturally
realized with the quantum states of the ring of metallic islands
(or atomic sites) connected by resonant tunnelling in the presence
of the Aharonov-Bohm flux threading the ring, a persistent-current
(PC) loop [19], and placed in an external electric field
perpendicular to the magnetic flux to perform the qu-gate manipulation
in the invariant subspace of two degenerate states. We focus on the
quantum mechanical aspects of qu-bit and qu-gate operations
with the PC loops leaving for future the discussion on
their practical implementation.\\

\section{Persistent-current qubit}
In the mesoscopic ring of a normal metal of size $L$ smaller than
the phase-decoherence length of the electrons, the charge current is
produced under the influence of Aharonov-Bohm (AB) flux [20].
Physically, the shifted energy minimum in the presence of AB flux
is counterbalanced by a net charge flow producing a persistent
current in the absence of resistive effects. This phenomenon was
predicted in [21], rediscovered in [22], and probed in an
experiment in [23-25]. The magnitude of persistent current in a
clean metallic ring is typically given by $J_c \sim ev_F/L$ where
$v_F$ is the electron Fermi velocity. In a nanoscopic (atomically
small) ring with one electron, the magnitude of the maximal persistent
current is $J_c\sim e\vert t_0\vert/\hbar N^2$ where $N$ is the
number of sites in a ring and $t_0$ is the electron hopping
amplitude between the sites. The PC is created individually by
single electrons hence the fundamental flux quantum $\Phi_0=hc/e$
is twice larger than the Abrikosov or Josephson flux quantum
$\Phi_1=hc/2e$. This very fact may permit new effects to arise
when a single Josephson vortex or Abrikosov fluxon is used to
manipulate the normal flux in a PC ring. Particularly, it is
known that at $\Phi=\Phi_1$ the PC ring has a degenerate ground
state configuration which produces
a frustrated (entangled) quantum state. Similar to the recently proposed
superconducting qubits manipulating macroscopic coherence, the qubit
in the persistent normal current ring is defined in this doubly
degenerate ground state configuration with the exception of the
auxhiliary level in our case. The Hamiltonian of the system is
\begin{equation}
H=-T\sum_{n=1}^{N}(a^+_na_{n+1}e^{i\alpha}+a^+_{n+1}a_ne^{-i\alpha})
\end{equation}
where $a^+_n$ is a fermionic operator creating (and $a_n$,
annihilating) electron at site ${\bf R}_n$ in a ring, $a_{N+1}=a_1$, and
$\alpha$ is the Aharonov-Bohm phase related to magnetic flux
threading the ring by
$\alpha=2\pi\Phi/N\Phi_0$. The Hamiltonian (1) is diagonalized by
the angular momentum (i.e., $m=0,1,\dots$) eigenstates $A^+_m\vert 0\rangle$ where
\begin{equation}
A^+_m=\frac{1}{\sqrt{N}}\sum_{n=1}^{N}e^{2\pi imn/N}a^+_n
\end{equation}
with the site energies
$\varepsilon_m=-2T\cos\frac{2\pi}{N}(m+\frac{\Phi}{\Phi_0})$
plotted against the normalized flux $\Phi/\Phi_0$ in Fig.1.
In our case here, the number of sites $N=3$.\\
Since two ground states are degenerate at $\Phi_0/2$, they can be
used as the components of qubit while the third one is coupling the
qubit to a qu-gate, to be discussed in the next two sections. The
practical realization of the qubit with desired architecture is
sketched in Fig.1. One possibility may be a three-sectional
normal-metal ring intersected by tunnelling barriers
(or consisting of overlapping metallic films separated by thin oxide layers).
Creating strong magnetic field to operate the qubit at the half quantum
flux is suggested with the help of superconducting fluxon trapped
in a hole inside the superconducting film, with lines of magnetic
field further focused by a
mesoscopic ferromagnetic cylinder near the ring.\\
The isolated qubit structure can in principle be realized as a
three-site defect in an insulating crystal, similar to
negative-ion triple vacancy (known as $F_3$-center) in the alkali
halide crystals (e.g., see [26]). The gate manipulation in such a
structure is provided by applying the electric field perpendicular
to the Aharonov-Bohm flux which will allow performing operations
in an invariant subspace of two degenerate states
(clockwise and counterclockwise persistent current directions).\\

\section{Qubit operations in an invariant subspace of one ring}
In the terms of operators $A^+_m$, ~the Hamiltonian (1) is
transformed into the diagonal form (we scale all energies in units of $T$)
\begin{equation}
H_0=\sum_{m}\varepsilon_mA^+_mA_m =
\pmatrix{-1 &0 &0\cr
          0 &2 &0\cr
          0 &0 &-1
          }.
\end{equation}
The interaction with an electrostatic potential $V_n=V_0\cos(2\pi n/3)$
at site $n$,
introduced with potential electrodes inclosing the ring (Fig.2), is presented
with the interaction term
\begin{equation}
H_1(V_0)=\frac{V_0}{2}\pmatrix{0 &1 &1\cr
                        1 &0 &1\cr
                        1 &1 &0
                       }.
\end{equation}
We also introduce, for further use, the interaction with a static site
potential $V_S$ (the Hamiltonian $H_2$), and the Hamiltonian $H_3$
representing the effect of the shifted magnetic flux away from $\Phi_0/2$
on the eigenenergy levels
$\Delta f =(\Phi-\Phi_0/2)/\Phi_0$
\begin{equation}
H_2 = V_S{\mbox{diag}}(1, 1, 1)  {\mbox{   and   }}
H_3 = {\mbox{diag}}(\Delta\varepsilon_1, \Delta\varepsilon_2,
\Delta\varepsilon_3)
\end{equation}
(remember that $V_0$, $V_S$ are in units of $T$). It is shown below
that the first two Hamiltonians
$H_0$ and $H_1$ suffice in the realization of the fundamental gates
except the phase shift. The $H_2$ and $H_3$ perform relative phase shift
between the qubit states.
The time dependence of the amplitudes $C_n(t)$ in the
angular momentum basis are found as
\begin{equation}
C_n(t) = \sum_{m}[\exp(-iHt)]_{mn}C_m(0)
\end{equation}
where, in general, $H=(H_0+H_1+H_2+H_3)$.
At this moment, we consider $V_S=0$ and
$\Phi=\Phi_0/2$ hence $H_2$ and $H_3$ do not exist.
When the interaction $H_1$ is turned on between $t=0$ and $t$ and zero otherwise,
the amplitudes are found by
\begin{equation}
C_n(t) = \sum_{m,k}S^{-1}_{kn}(V_0)e^{-iE_k t}S_{mk}(V_0)C_m(0)
\end{equation}
where $E_k(V_0)$ are the eigenenergies of Hamiltonian
$H_0+H_1(V_0)$ and $S_{nm}(V_0)$ are the unitary matrices transforming
from the noninteracting eigenbasis (the ones corresponding to
$H_0$) to the eigenbasis of the full Hamiltonian $H_0+H_1$.
Inspection of Eq.(7) shows that, at fixed values of $V_0$ and the
evolution time $t$, the population of the auxhiliary state (in
the noninteracting basis) vanishes if it was initially set to zero.
This happens when the three energies $\varepsilon_n(V_0)$, ($n=1,2,3$) are
commensurate, so that
exponential factors in Eq.(7) destructively interfere with each other at
fixed instants in time.\\
The eigenenergies $E_k(V_0)$ are plotted in Fig.3.
The amplitudes vanish when the commensurability condition
\begin{equation}
E_3-E_1 = K\,(E_2 -E_3)
\end{equation}
is satisfied for integer $K$. The corresponding values of the potential
that does Eq.(8) are
\begin{equation}
V_0(K)=-\frac{2}{3K}[K^2+K+1+(K-1)\sqrt{K^2+4K+1}].
\end{equation}
In particular we mention that for $K=1$ one has $V_0^{(1)}=-2$; and at $K=3$
one has
~$V_0^{(3)}=-\frac{2}{9}(13+2\sqrt{22})=-4.9735$. As shown below these two
cases give the bit-flip and Hadamard transformations. The $K=1$ case can be
explicitly proved by checking the identity
\begin{equation}
\exp\{-it\
\pmatrix{-1 &-1 &-1\cr
         -1 &2  &-1\cr
         -1 &-1 &-1
         }
         \} =
\frac{1}{2}\pmatrix{1+c+s  &s        &-1+c+s\cr
                    s      &2(c-s)   &s\cr
                    -1+c+s &s        &1+c+s
         }
\end{equation}
where $c=\cos(t\sqrt{6})$, $s=i\sqrt{\frac{2}{3}}\sin(t\sqrt{6})$.
At $s=0$ (i.e. $c=1$), the transformation matrix of Eq.(10)
block-diagonalizes in a subspace of states 1,3 (i.e. $\vert 0\rangle$, $\vert 1\rangle$)
and the upper state 2 (i.e. $\vert c\rangle$), after reshuffling state numbering
from (1,2,3) to (1,3,2).\\
In Fig.4 the populations of the states $p_n(t)=\vert C_n(t)\vert^2$ are
plotted for these two cases $K=1$ and $K=3$. At times $t_1$ for
$K=1$ and $t_3$ for $K=3$, the population of the auxhiliary state
vanishes assuming that it was zero at $t=0$. These special times are
(in units of $\hbar/T$)
\begin{equation}
t_1 = \frac{\pi}{\sqrt{6}} = 1.2825, \quad t_3=
\frac{\pi}{2[\varepsilon_2(V)-\varepsilon_3(V)]_{V=V_0(3)}} =
0.7043
\end{equation}
where
\begin{equation}
\varepsilon_{1,3}(V)=\frac{1+V/2}{2}\mp\frac{3}{2}\sqrt{1-V/2+V^2/4},
\quad \varepsilon_2(V)=-1-V/2
\end{equation}
for $V \le 0$. We notice that
the configuration $(t_1, K=1)$ performes the bit-flip
$\vert 0\rangle ~~\leftrightarrow~~\vert 1\rangle$
whereas $(t_3, K=3)$ creates equally populated Hadamard-like
superpositions of $\vert 0\rangle$ and $\vert 1\rangle$. These
operations are presented in the qubit subspace by the matrices
(overall phases are not shown)
\begin{equation}
G_1 = \pmatrix{0  &1\cr
         1 &0
         }  {\mbox{  and  }}
G_3=\frac{1}{\sqrt{2}}
\pmatrix{1   &-i\cr
         -i  &1
         }.
\end{equation}
The $G_1$ gate manifests the bit-flip whereas $G_3$ is different
from the standard Hadamard by a relative $\pi/2$ phase.
The relative phase in $G_3$ can be changed by an additional procedure
using the diagonal Hamiltonians $H_2$ and $H_3$ defined above by introducing
additional relative energy difference between the qubit states without
changing the couplings and the eigenstates (note that $H_{0}$, $H_{2}$
and $H_{3}$ are simultaneously diagonal). If the operational space is a
double-qubit space
the overall phase of the qubit states becomes important. For instance,
the overall qubit phase can be corrected with the unitary matrix
$\exp(-i(H_0+H_2)t)$ at a fixed and determinable
configuration $(t^{*},V_0^{(*)})$
whereas the relative phase between the qubit states
can be changed using the phase rotation matrix
\begin{equation}
G_2(\phi) = \pmatrix{e^{i\, \phi}  &0\cr
         0 &e^{-i\, \phi}
         }
\end{equation}
in the form of an Euler-type transformation
$G_{2}(-\pi/4)\,G_3\,G_{2}(-\pi/4)$. The fixed phase value $-\pi/4$
can be obtained by turning off $H_1$ and $H_2$ and turning on $H_3$
[i.e. $H=(H_0,0,0,H_3)$] for the required time. Since both Hamiltonians are
diagonal, the qubit subspace is invariant under this transformations for all
evolution times.\\
The gate operations described in this section can be regarded as the
(quenched) Rabi oscillations
affected by the nondiagonal matrix elements generated by $H_1$.
The transitions between the degenerate states are
achieved through virtual transitions to an auxhiliary eigenstate
$\vert c\rangle$ with a sufficiently higher energy level
(Fig.5). Switching off the interaction, when the auxihiliary state becomes
depopulated, brings the final configuration back to the qubit subspace.\\

\section{Quantum computing procedures with one and two persistent-current
rings} The standard procedures of quantum computation are the
initialization (input), the logic gate transformations in one ring,
the controlled bit flips on the qubit pairs
(the {\em CNOT}), and the reading of the output to a classical device.
We discussed the bit-flip, Hadamard as well as the phase shift above.
Here, we discuss the initialization, the two-qubit CNOT gate and the
read-out operations. \\

(a){\em Initialization}.
Adiabatically shift the
magnetic flux in each ring from half flux quantum
and allow the system to relax to the nondegenerate lowest energy state
$\vert 0\rangle$ by spontaneous emission. By applying
$G_3$, we receive a state of equally superposed degenerate levels which
is conventionally the  initial state in some quantum computing algorithms,
in particular Shor's algorithm for factorizing large integers [1].\\

(b){\em CNOT}. This gate performs a conditional bit-flip in the
qubit No.2 when the qubit No.1 is in a fixed state. To make such a
transformation a non-demolition measurement of the state of the
first qubit is needed. We use the direction of the current in the
first qubit to control the bit flip in the second qubit. The
suggested scheme is sketched in Fig.6. The flux in the qubit No.1
(which includes the externally applied flux and the flux created
by a persistent current) is extracted from the former by a
$-\Phi_0/2$ compensating coil, and further supplied to the Hall
bar with a (large) current passing through it. The Hall voltage
generated in the bar, after extraction of the voltage
corresponding to fixed value related to the counterclockwise
current direction and applied to the $V$ electrodes of qubit No.2
produces, if the voltage proves nonzero, the flip of the second
qubit. The procedure may in principle be executed in a totally
reversible way if the Quantum Hall Effect regime is manifest [29].\\
In more detail, assume that the current generated by a gubit loop, ($J_Q$),
is $J_Q=J_0$ when the qubit is in the clockwise direction of rotation
(a state $\vert 0\rangle$), and value of the current $J_Q=-J_0$ when the direction
of rotation is counterclockwise (i.e., qubit is in a state $\vert 1\rangle$).
Supply a current thus received from the first qubit, with a fixed value
$J_0$ added, $J_Q+J_0$, to the Hall bar which will produce
a voltage $VQ=k*(J_Q+J_0)$. Applying this one to the qubit No.2 will produce
a voltage $2kJ_0=-2$ (in units of $T$) at a proper choice of the
instrumental parameter $k$
(i.e., the magnitude and the sign of the transport current in a bar $J$).
The voltage is nonzero when qubit is $\vert 0\rangle$,
and zero when it is in a state $\vert 1\rangle$.
Therefore, when the voltage output from qubit No.1 is applied to the qubit
No.2 at the time
interval $t_1$ which is specified in Eq.(11), the qubit No.2 will flip its
state or
remain idle depending on whether the qubit No.1 is in the $\vert 0\rangle$ or
in the $\vert 1\rangle$ states, i.e. the $\em CNOT$ operation.\\

(c){\em Reading the output}. Reading the result, i.e. the
population of the qubits when computation concludes
 can be realized with the same device as one shown in
Fig.6 by measuring the voltage on a Hall probe since it is
directly proportional to the persistent current, and therefore, by its
polarity, distinguishes between the clockwise and the
counterclockwise directed currents.
The device is analogous to a
Stern-Gerlach sensor of Ref.[1], or to the Ramsey-zone measuring
devices of the optical beam polarization qubits in Ref.[3].\\

\section{Conclusions}
In conclusion, we presented a new method, using the
normal-metal (non-superconducting) persistent-current states [21]
in three-site resonantly coupled quantum dots (or in the triple anionic
vacancy sites), in perpendicular magnetic and electric fields.
The operations of the qu-gate are similar to the previous mechanisms within the
qubit subspace, but the presence of the auxhiliary level manifests the
quantum gates without using external switches and contacts.
The particular persistent current realization of this three level system
however may be hard in practice for
many reasons including the necessity of extremely
high magnetic field (in the case of a nanoscopic variant with triple
vacancies), or very large current-to-voltage conversion ratios for
${\em CNOT}$ transformation (in case of a mesoscopic variant with
triples of quantum dots); the decoherence and the instrumental errors.
However, it is very likely that the simplicity of the theoretical mechanism
may permit other manifestations to exist.
This part is not discussed in the present work in any practical detail
except for mentioning that use of high-$\mu$, high-$\varepsilon$
materials may resolve some of the problems (see in this
respect a recent work [27] in which the Aharonov-Bohm oscillations
have been observed in a ring with permalloy). Another trend may be
in using of the quantum effects other than the Aharonov-Bohm
mechanism (the Berry phase [28], the spin-orbit coupling, etc.)
for creating other persistent current structures similar to one
studied in present work.\\
The suggested qubit differs from those currently
investigated by the quantum-computing community (the photon beams,
trapped ions, liquid-state NMR atoms, superconducting Cooper-pair
islands or $\pi$-Josephson junctions) in the respect of the
auxiliary register $\vert c\rangle$ which provides a
possibility of coherent coupling of its operational $\vert 0\rangle$ and
$\vert 1\rangle$ registers to (or the mere execution of) the logical
reversible {\em NOT} and {\em CNOT} gates. Unlike the
superconducting (macroscopic) qubits which may be shown to display
strong decoherence during switching between the states of opposite
direction superconducting Josephson currents (a phase-slip
mechanism), the persistent-current qu-bit performs similar
transformation (the qu-gate) as a process intrinsically
incorporated into their three-level structure and hence not
involving the additional decoherence.\\
The persistent-current scheme is flexible for
further modifications including mechanisms of error correction
(not discussed in present work),
the addition of extra registers for more complex unified bit-gate
manipulation, and is expected to better fit the requirements of keeping
quantum coherence within the operational cycle of the computer by
addressing the memory and the processor registers
in the same quantum unit.\\

\section{Acknowledgements}
One of authors (I.O.K.) acknowledges hospitality of Department of Physics,
University of Naples ``Federico II'' for part of work on subject of the paper.\\


\newpage

\section*{Figure Captions}

Fig.1:(a)A sketch of the magnetically focused lines of the magnetic
field from the superconducting fluxon trapped in the opening of
superconducting foil ($S$), compressed by ferromagnetic crystal
($F$) and directed into the interior of PC ring ($R$);
(b)Operational diagram of the qubit at half-flux quantum,
$\Phi=\Phi_0/2$. 1,3 - degenerate states carrying opposite
currents $\pm j=-c\partial\varepsilon/\partial\Phi$ (full
and dashed lines), 2 - zero-current virtual state (a control
state, dotted line) which couples the qubit to the logical qu-gate.\\

Fig.2: A sketch of the bit flip. $C$ is an output
coil, $V$'s are the electrodes creating electric field perpendicular
to the magnetic field (normal to sheet) in a qubit.\\

Fig.3: Energy versus electrostatic potential. 1 and 3 (solid line
and dotted line) are the energies which become degenerate at $V_0=0$, and
2 (the dashed line) is an energy of the auxiliary control state
$\vert c\rangle$. The arrows indicate the values of the potential $V_0$ 
corresponding to the operational points of the bit-flip (i.e. $G_1$) and 
$G_3$ (i.e. Hadamard) gates. \\ 

Fig.4: Evolution diagrams of quantum gates $G_1$ (panel ${\em a}$)
and $G_3$ (panel ${\em b}$).
Solid and dashed lines show the time dependence of the
population of the states $\vert 0\rangle$ and $\vert 1\rangle$ which are 
degenerate
at $V_0=0$. The dotted lines show the time dependence of the auxiliary
population. The arrows indicate the ``operational point'' of the qu-gate,
the time of evolution corresponding to the return to the invariant qubit 
subspace.\\

Fig.5: Level diagram of the persistent-current qubit. Arrows indicate the
virtual transition to the auxiliary state at the fixed-time
interval (quenched) Rabi oscillation.\\

Fig.6: The sketch of a possible physical implementation of the
{\em CNOT} logical gate. $C$ - current-response loop of a control
qubit (qubit No.1), $F$ - flux-$\Phi_0/2$ compensation loop, $H$ -
Hall-effect sensor generating the potential $V_0$ controlling qubit
No.2. $J_0$ is a fixed current which should be large enough to
create a strong voltage output signal ensuring effective control
over the second qubit.\\

\newpage

\begin{figure}
\hspace{10mm}
\psfig{figure=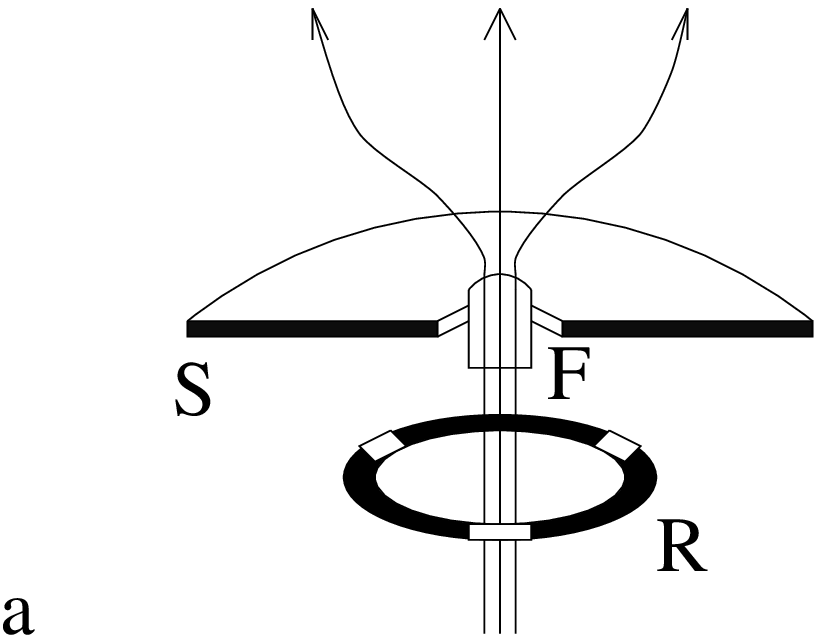,width=12cm,height=10cm}
\end{figure}
\vspace{5mm}
\begin{center}
Fig.1a
\end{center}
\newpage

\begin{figure}
\hspace{10mm}
\psfig{figure=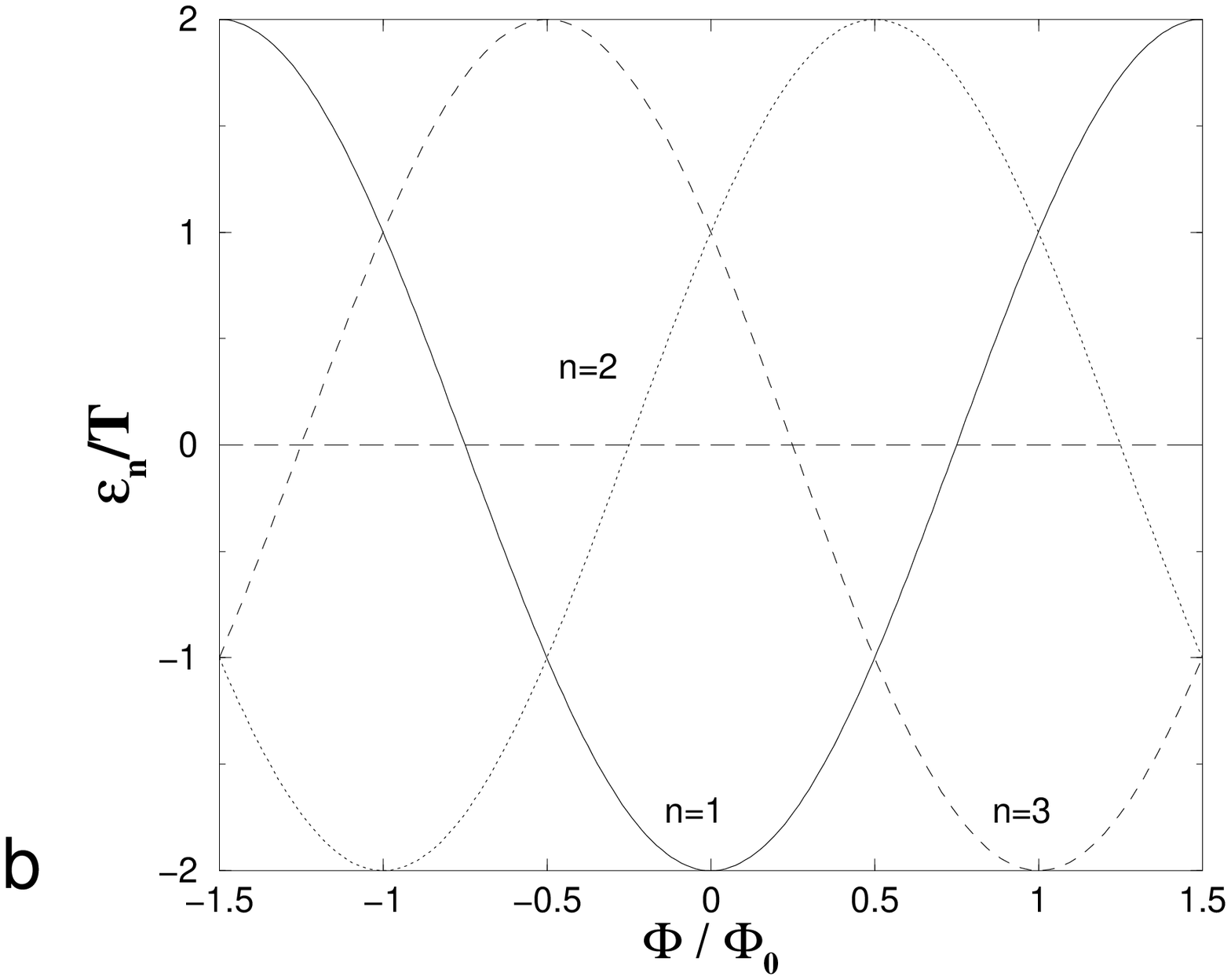,width=12cm,height=8cm}
\end{figure}
\vspace{5mm}
\begin{center}
Fig.1b
\end{center}
\newpage

\begin{figure}
\hspace{10mm}
\psfig{figure=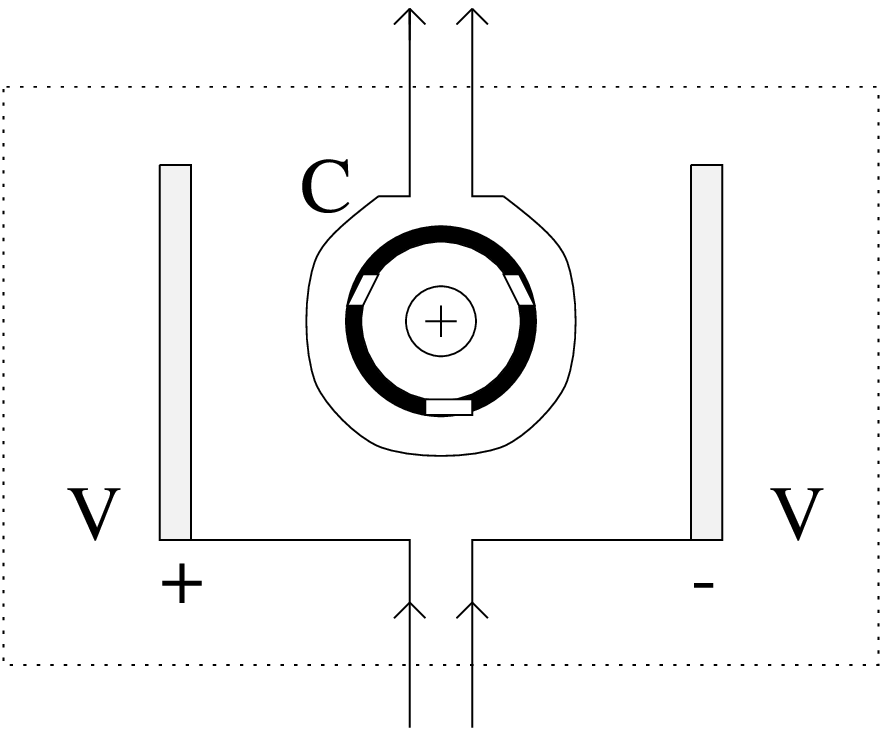,width=10cm,height=8cm}
\end{figure}
\vspace{5mm}
\begin{center}
Fig.2
\end{center}
\newpage

\begin{figure}
\hspace{10mm} \psfig{figure=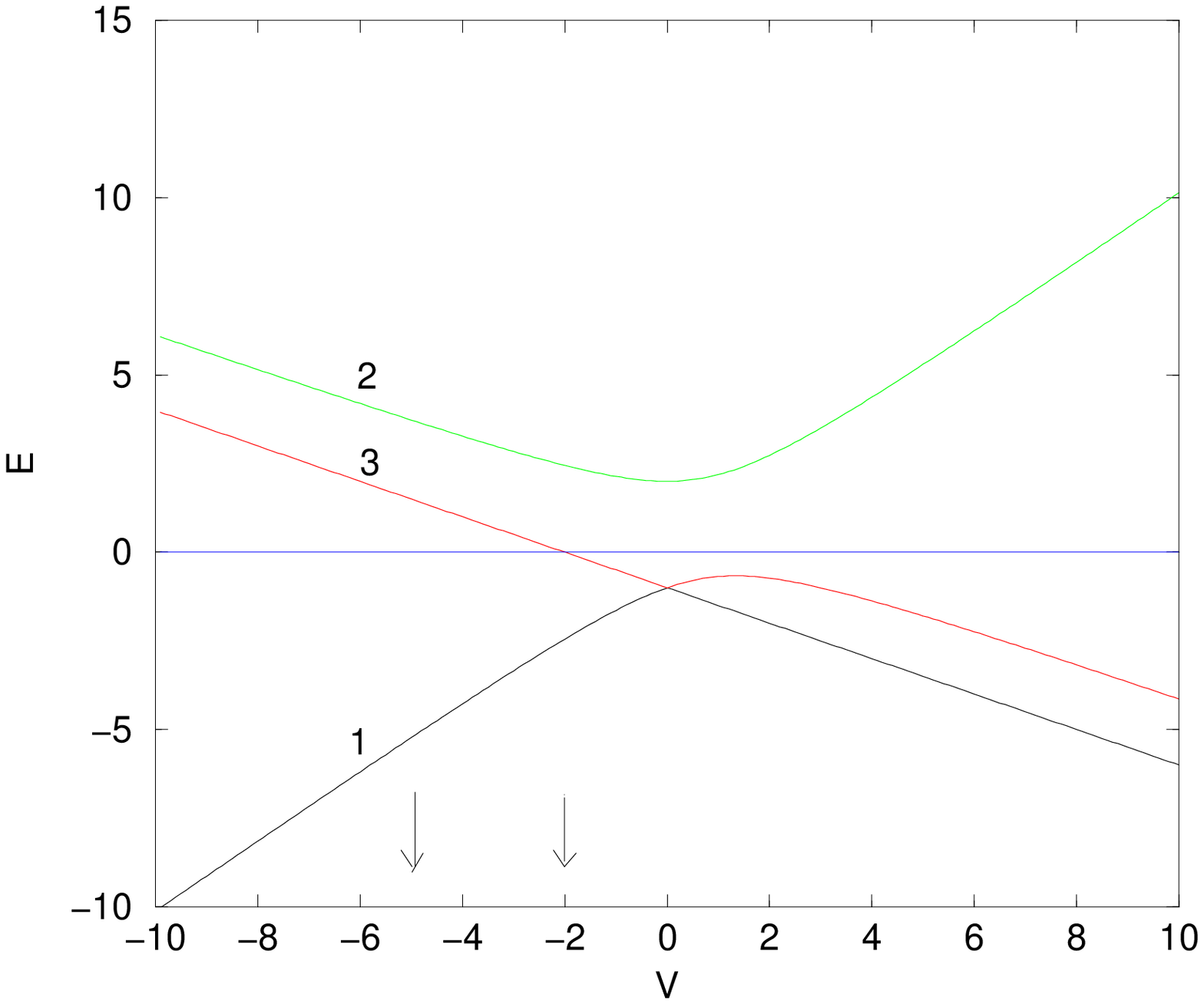,width=10cm,height=7cm}
\end{figure}
\vspace{5mm}
\begin{center}
Fig.3
\end{center}
\newpage

\begin{figure}
\hspace{10mm} \psfig{figure=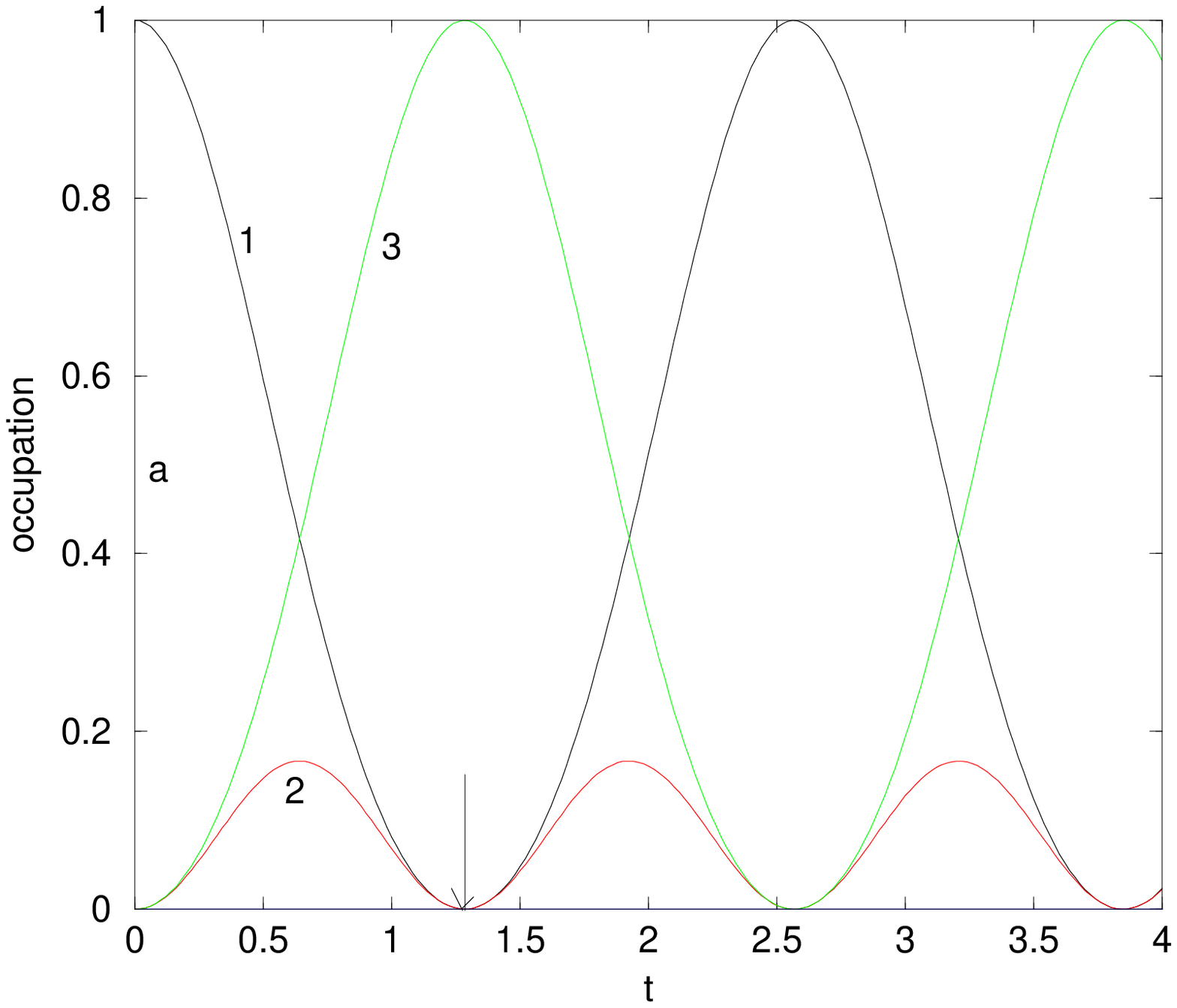,width=12cm,height=8cm}
\end{figure}
\vspace{5mm}
\begin{center}
Fig.4a
\end{center}
\newpage

\begin{figure}
\hspace{10mm} \psfig{figure=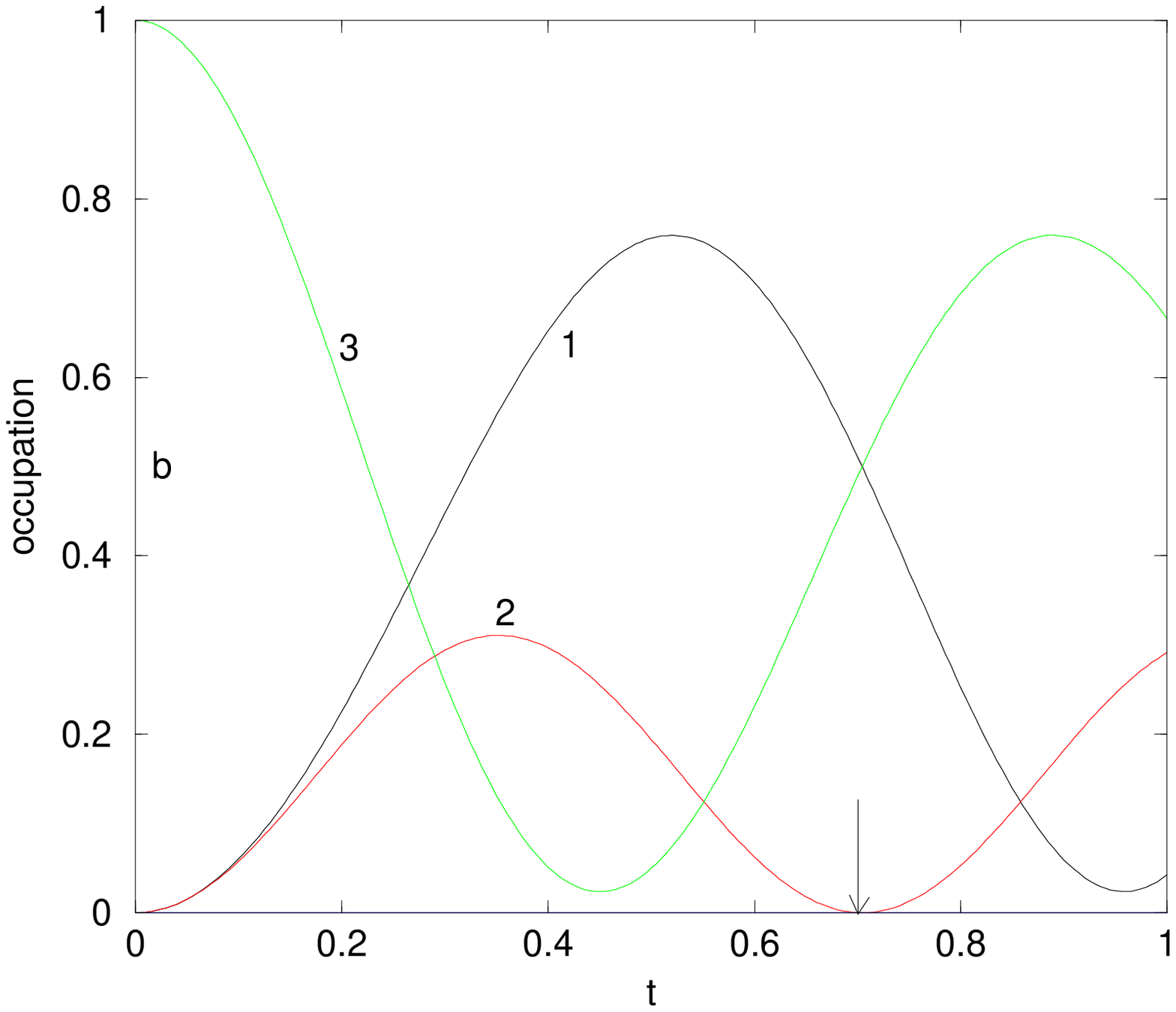,width=12cm,height=8cm}
\end{figure}
\vspace{5mm}
\begin{center}
Fig.4b
\end{center}
\newpage

\begin{figure}
\hspace{20mm} \psfig{figure=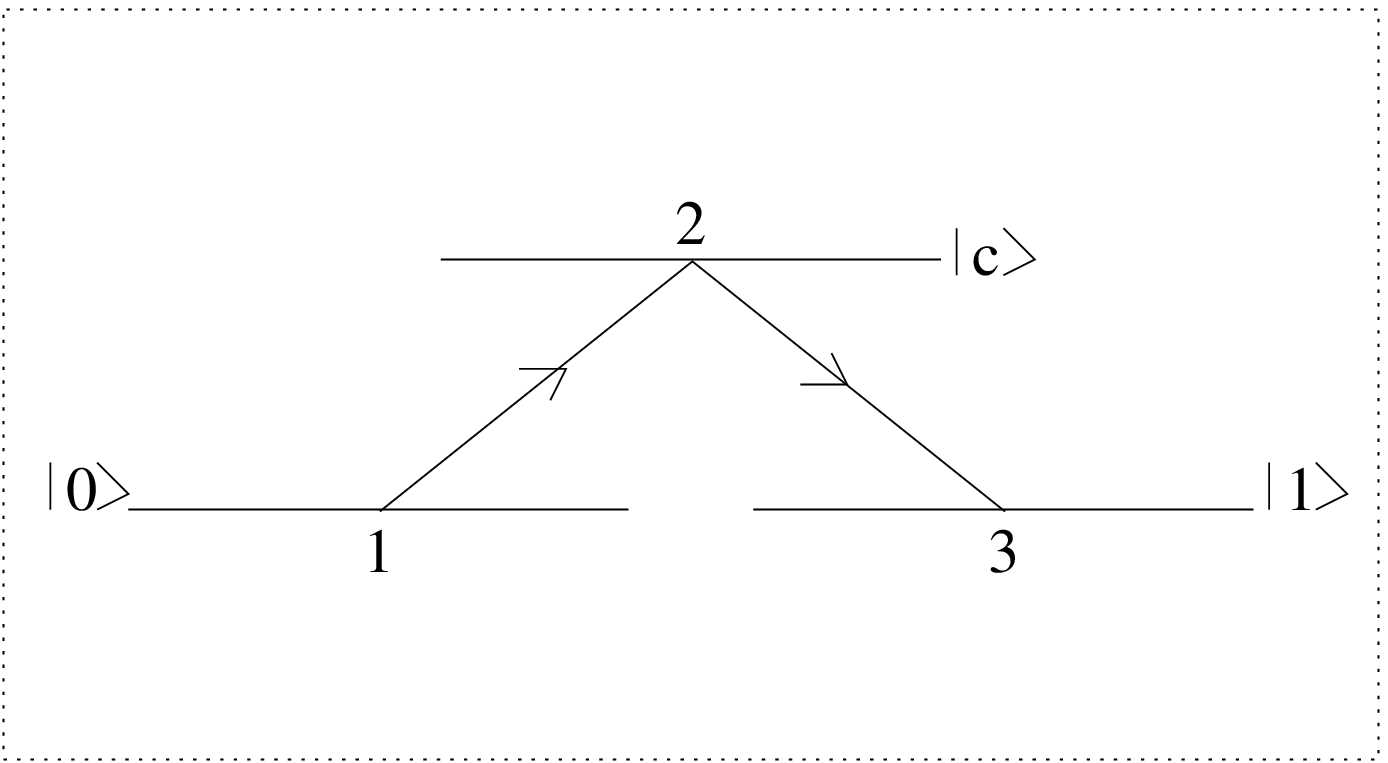,width=10cm,height=5cm}
\end{figure}
\vspace{5mm}
\begin{center}
Fig.5
\end{center}
\newpage

\begin{figure}
\hspace{10mm} \psfig{figure=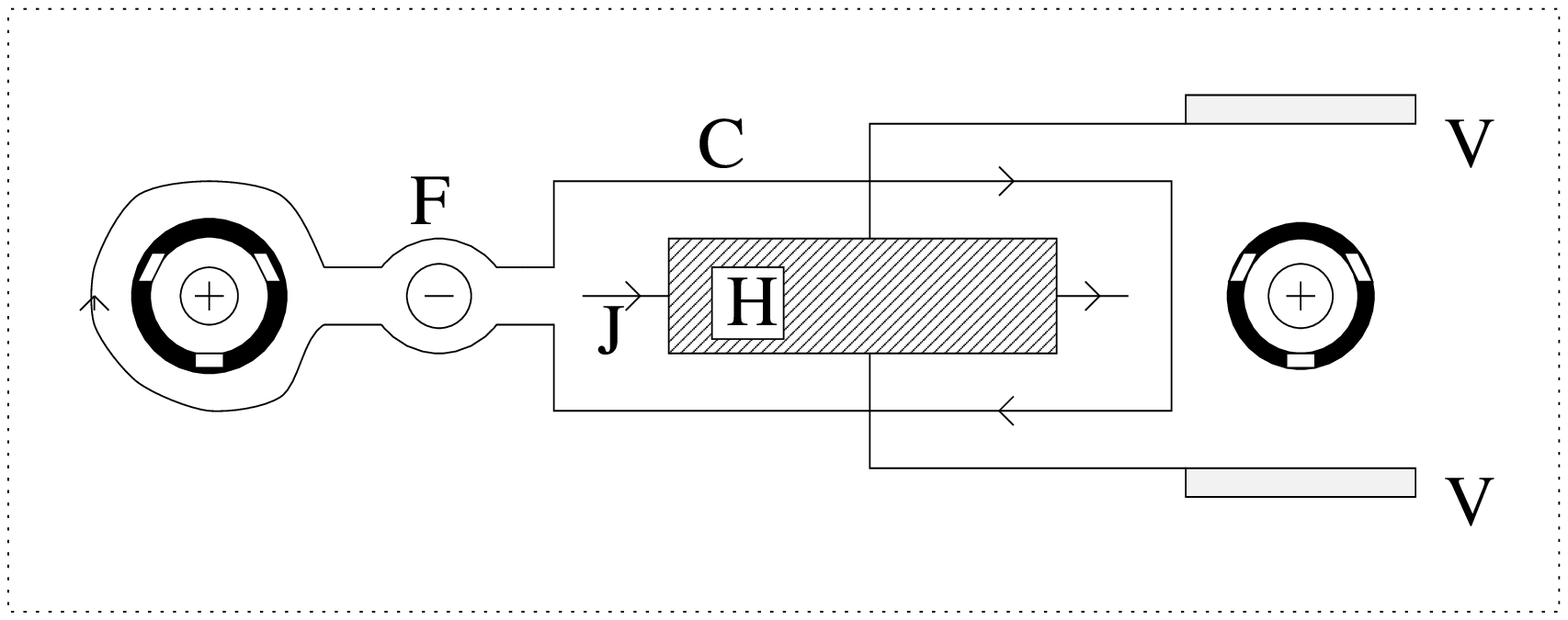,width=14cm,height=6cm}
\end{figure}
\vspace{5mm}
\begin{center}
Fig.6
\end{center}


\begin{thebibliography}{99}
\bibitem{ref1}A. Barenco, Contemp. Phys. {\bf 37}, 375 (1996).
\bibitem{ref2}A. Steane, Rep. Progr. Phys. {\bf 61} , 117 (1998).
\bibitem{ref3}C. H. Bennett, Phys. Today, p.24, Oct.1995.
\bibitem{ref4}D. Deutsch, Proc. Roy. Soc. London, {\bf A 400}, 97 (1985).
\bibitem{ref5}V. Vedral, A. Barenco, and A. Ekert, Phys. Rev. {\bf A54}, 147 (1996).
\bibitem{ref6}P. W. Shor, in: Proc. 35th Annual Symp. Th. Comp. Science, p.124.
Ed. S. Goldwasser, IEEE Comp. Soc. Press, Los Alamos, 1994.
\bibitem{ref7}D. Loss and D. P. DiVincenzo, Phys. Rev. {\bf A57}, 120 (1998).
\bibitem{ref8}Y. Nakamura, C. D. Chen, and J. S. Tsai, Phys. Rev. Lett. {\bf 79},
2328 (1997).
\bibitem{ref9}A. Shnirman, G. Sch\"{o}n, and Z.Hermon, Phys. Rev. Lett,
{\bf 79}, 2371 (1997).
\bibitem{ref10}Y. Maknin, G, Sch\"{o}n, and A. Shnirman, Nature {\bf 398}, 305
(1999).
\bibitem{ref11}V. Bouchiat, D. Vion, P. Joyez, D. Esteve, and M. H. Devoret,
Phys. Scripta {\bf T76}, 165 (1998).
\bibitem{ref12}T. P. Orlando, J. E. Mooij, Lin Tian, C. H. van der Wal,
L. S. Levitov, S.Lloyd, and J. J. Mazo, Phys. Rev.{\bf B60}, 15398 (1999);
Science, {\bf 285}, 1036 (1999).
\bibitem{ref13}L. V. Ioffe, V. B. Geshkenbein, M. V. Feigelman, A. L. Fauchere,
and G. Blatter, Nature {\bf 398}, 679 (1999).
\bibitem{ref14}A. M. Zagoskin, ``A scalable, tunable qubit, based on a clean
DND or grain boundary D-D junction'', preprint cond-mat/9903170.
\bibitem{ref15}D. D. P. Di Vincenzo, G. Burkard, D. Loss, and E. V. Sukhorukov,
``Quantum computation and spin electronics'', in: Quantum Mesoscopic
Phenomena and Mesoscopic Devices in Microelectronics, p.399. Eds. I. O. Kulik
and R. Ellialtioglu. Kluwer,
Dordrecht, 2000.
\bibitem{ref16}Y. Makhlin, G. Sch\"{o}n, and A. Shnirman,
Rev. Mod. Phys., {\bf 73}, 357 (2001).
\bibitem{ref17} D. V. Averin, Solid St. Commun. {\bf 105}, 659 (1998).
\bibitem{ref18} D. Bacon, J. Kempe, D. A. Lidar, and K. B. Whaley,
Phys. Rev. Lett. {\bf 85}, 1758 (2000);
A. Beige, D. Braun, B. Tregenna, and P. L. Knight,
Phys. Rev. Lett. {\bf 85}, 1762 (2000).
\bibitem{ref19}I. O. Kulik, ``Non-decaying currents in normal metals'', in:
{\it{ibid}}., Ref.[15], p.259.
\bibitem{ref20}Y. Aharonov and D. Bohm, Phys. Rev. {\bf 115}, 485 (1959).
\bibitem{ref21}I. O. Kulik, JETP Lett., {\bf 11}, 275 (1970).
\bibitem{ref22}M. Buttiker, Y. Imry and R. Landauer. Phys. Lett., {\bf A96}, 365
(1983).
\bibitem{ref23}V. Chandrasekhar, R. A. Webb, M. J. Brady, M. B. Ketchen, W. J.
Gallagher and A. Kleinsasser, Phys. Rev. Lett. {\bf 67}, 3578 (1991).
\bibitem{ref24}D. Mally, C. Chapelier and A. Benoit, Phys. Rev. Lett. {\bf 70},
2020 (1993).
\bibitem{ref25}E. M. Q. Jariwala, P. Mohanty, M. B. Ketchen and R. A. Webb,
Diamagnetic persistent current in diffusive normal-metal rings (to appear in
Phys. Rev. Lett.).
\bibitem{ref26}C. Kittel, Introduction to Solid State Physics. J. Wiley, New
York, 1996.
\bibitem{ref27}K. Tsubaki, Jpn. J. Appl. Phys. {\bf 40}, 1902 (2001).
\bibitem{ref28}M. V. Berry, Proc. Roy. Soc. (London) {\bf A392}, 45 (1984).
\bibitem{ref29}Perspectives in Quantum Hall Effects,
S. Das Sarma and A. Pinczuk, eds. Wiley, 1996.
\end{thebibliography}
\end{document}